\documentclass[conference]{IEEEtran}
\IEEEoverridecommandlockouts

\usepackage[normalem]{ulem}
\useunder{\uline}{\ul}{}
\usepackage{cite}
\usepackage{amsmath,amssymb,amsfonts}
\usepackage{algorithmic}
\usepackage{algorithm}  
\usepackage{graphicx}
\usepackage{textcomp}
\usepackage{xcolor}
\usepackage[perpage]{footmisc}
\usepackage{float} 
\usepackage{bm}
\usepackage{makecell}
\usepackage{multirow}
\usepackage{array}
\usepackage{booktabs}
\usepackage{multirow}
\usepackage{fancyhdr} 
\usepackage{color}
\usepackage{tabu}
\usepackage{ulem}
\usepackage{nomencl}
\makenomenclature
\usepackage{etoolbox}
\usepackage{graphicx}
\usepackage{subfigure}
\usepackage[cal=boondoxo]{mathalfa}
\newcolumntype{I}{!{\vrule width 1.3pt}}
\newlength\savedwidth

\newlength\savewidth

\def\BibTeX{{\rm B\kern-.05em{\sc i\kern-.025em b}\kern-.08em
    T\kern-.1667em\lower.7ex\hbox{E}\kern-.125emX}}

\begin{document}

\title{Machine Learning-enabled Traffic Steering in O-RAN: A Case Study on Hierarchical Learning Approach \\}

\author{Md~Arafat~Habib,
        Hao~Zhou,
        Pedro~Enrique~Iturria-Rivera, Yigit~Ozcan,
        Medhat~Elsayed,
        Majid~Bavand,\\
        Raimundas~Gaigalas,
        and~Melike~Erol-Kantarci,~\IEEEmembership{Senior Member, IEEE}
        \vspace{-20pt}
\thanks{Md~Arafat~Habib, Hao Zhou, Pedro~Enrique~Iturria-Rivera, and Melike Erol-Kantarci are with the School of Electrical Engineering and Computer Science, University of Ottawa, Ottawa, ON K1N 6N5, Canada (\{mhabi050, hzhou098, pitur008, melike.erolkantarci\}@uottawa.ca,).}
\thanks{Medhat Elsayed, Majid Bavand, Raimundas Gaigalas and Yigit Ozcan are with the Ericsson (e-mail:\{medhat.elsayed, majid.bavand, raimundas.gaigalas, yigit.ozcan\}@ericsson.com).
}}

\maketitle

\thispagestyle{fancy}   
\fancyhead{}                
\lhead{The paper has been accepted for publication in IEEE Communications Magazine. Personal use of this material is permitted. Permission from IEEE must be obtained for all other uses. \copyright2024 IEEE}
\cfoot{}
\renewcommand{\headrulewidth}{0pt} 

\begin{abstract}
Traffic Steering is a crucial technology for wireless networks, and multiple efforts have been put into developing efficient Machine Learning (ML)-enabled traffic steering schemes for Open Radio Access Networks (O-RAN). Given the swift emergence of novel ML techniques, conducting a timely survey that comprehensively examines the ML-based traffic steering schemes in O-RAN is critical. In this article, we provide such a survey along with a case study of hierarchical learning-enabled traffic steering in O-RAN. In particular, we first introduce the background of traffic steering in O-RAN and overview relevant state-of-the-art ML techniques and their applications. Then, we analyze the compatibility of the hierarchical learning framework in O-RAN and further propose a Hierarchical Deep-Q-Learning (h-DQN) framework for traffic steering. Compared to existing works, which focus on single-layer architecture with standalone agents, h-DQN decomposes the traffic steering problem into a bi-level architecture with hierarchical intelligence. The meta-controller makes long-term and high-level policies, while the controller executes instant traffic steering actions under high-level policies. Finally, the case study shows that the hierarchical learning approach can provide significant performance improvements over the baseline algorithms.
\end{abstract}


\vspace{-5pt}
\section{Introduction}

5G beyond and envisioned 6G networks are expected to accommodate diverse use cases at a large scale, which requires automated control and optimization of network functionalities. However, the existing cellular architecture lacks the ability to provide precise control over the Radio Access Network (RAN) at a granular level Open RAN paradigm introduces an open architecture that facilitates closed-loop control, data-driven decision-making, and intelligent optimization of the RAN \cite{2}. The radio controller within an O-RAN compliant (an implementation of Open RAN paradigm \cite{21}) architecture can be divided into two main components: the Near-Real-Time RAN Intelligent Controller (near-RT-RIC) and the Non-Real-Time RAN Intelligent Controller (non-RT-RIC) \cite{2}. Positioned at the top of the hierarchy, the non-RT-RIC supports rApps that execute high-level RAN optimization tasks. Non-RT-RIC has access to network information and offers AI-enabled insights to the near-RT-RIC. On the other hand, the near-RT-RIC, located at a lower level, facilitates control and optimization of RAN elements by employing RAN applications referred to as xApps in the O-RAN terminology. Such disaggregated RAN schemes exhibit a programmable and highly modular architecture, making them well-suited for developing advanced AI-based modules that optimize networks via rApps and xApps. 

Traffic Steering is one of the prime use cases of O-RAN \cite{4}. To design an efficient traffic steering xApp for O-RAN, the RIC faces the challenge of managing multiple combinations of Radio Access Technologies (RATs) and traffic types to meet stringent quality-of-service requirements \cite{5}. Furthermore, developing a robust traffic steering scheme for O-RAN can get more complicated since 5G networks encompass densely deployed Base Stations (BSs) with many users having diverse traffic profiles. 
To this end, Machine Learning (ML) techniques especially Reinforcement Learning (RL) have been considered as ideal solutions to handle such highly dynamic environments. For instance, Lacava et al. propose a traffic steering scheme based on conservative Q-learning for user-specific traffic steering in O-RAN \cite{1}. Tamim et al. propose a DQN-based approach that involves predicting network congestion and taking preemptive measures to perform traffic steering to minimize anticipated queueing delays \cite{6}. However, DQN may fail to achieve faster convergence in more challenging RL problems formulated for O-RAN-based traffic steering, which may hamper the efficiency of real-time systems.

With inspiration for developing more efficient traffic steering solutions, this work first provides background on adapting traffic steering in O-RAN. Next, we comprehensively review state-of-the-art ML algorithms for traffic steering, including supervised and unsupervised learning, RL, federated learning, graph learning, and so on. Furthermore, we introduce a hierarchical learning-based traffic steering scheme for O-RAN that unlocks hierarchical intelligence. In particular, we use hierarchical DQN (h-DQN), which applies a bi-level architecture to decompose the traffic steering problem in O-RAN. Decomposing the traffic steering problem using h-DQN can bring higher exploration efficiency, faster convergence, and better network performance\cite{12,14}. The hierarchical structure of h-DQN allows for flexible decision-making at two distinct levels of the network hierarchy via two different agents interacting with the environment. Agents at lower levels can monitor and make local decisions, such as steering traffic to specific BSs. Higher-level agents can oversee the overall network and adjust global traffic steering strategies based on network-wide performance metrics.           

In this paper, we provide an extensive survey on ML algorithms and their application in traffic steering, marking the first comprehensive exploration in this area to our knowledge. Our primary contribution is the introduction of a novel hierarchical learning scheme specifically tailored for traffic steering in O-RAN. This innovative approach represents a significant advancement in the field. Additionally, we present a detailed case study demonstrating the practical implementation of this scheme through an example of Hierarchical Reinforcement Learning (HRL), showcasing its effectiveness and potential in real-world scenarios. This paper not only surveys existing methodologies but significantly extends the current understanding of intelligent traffic steering in O-RAN through our unique hierarchical learning proposition. Unlike our previous works \cite{3,14} on traffic steering where we had proposed standalone machine learning algorithms to perform traffic steering in a multi-RAT scenario, this paper focuses on providing a road map to implement hierarchical learning in O-RAN that includes a case study as a proof of concept.      

The rest of the paper is organized as follows. Section \ref{s2} provides background on traffic steering in O-RAN and Section \ref{s3} surveys relevant ML techniques. Section \ref{s4} elaborates on hierarchical learning for traffic steering and presents the h-DQN implementation as a case study. Section \ref{s5} presents the case study results, and Section \ref{s6} concludes this work. 

\begin{figure}[!t]
\setlength{\abovecaptionskip}{-5pt}
\centerline{\includegraphics[width=0.6\linewidth]{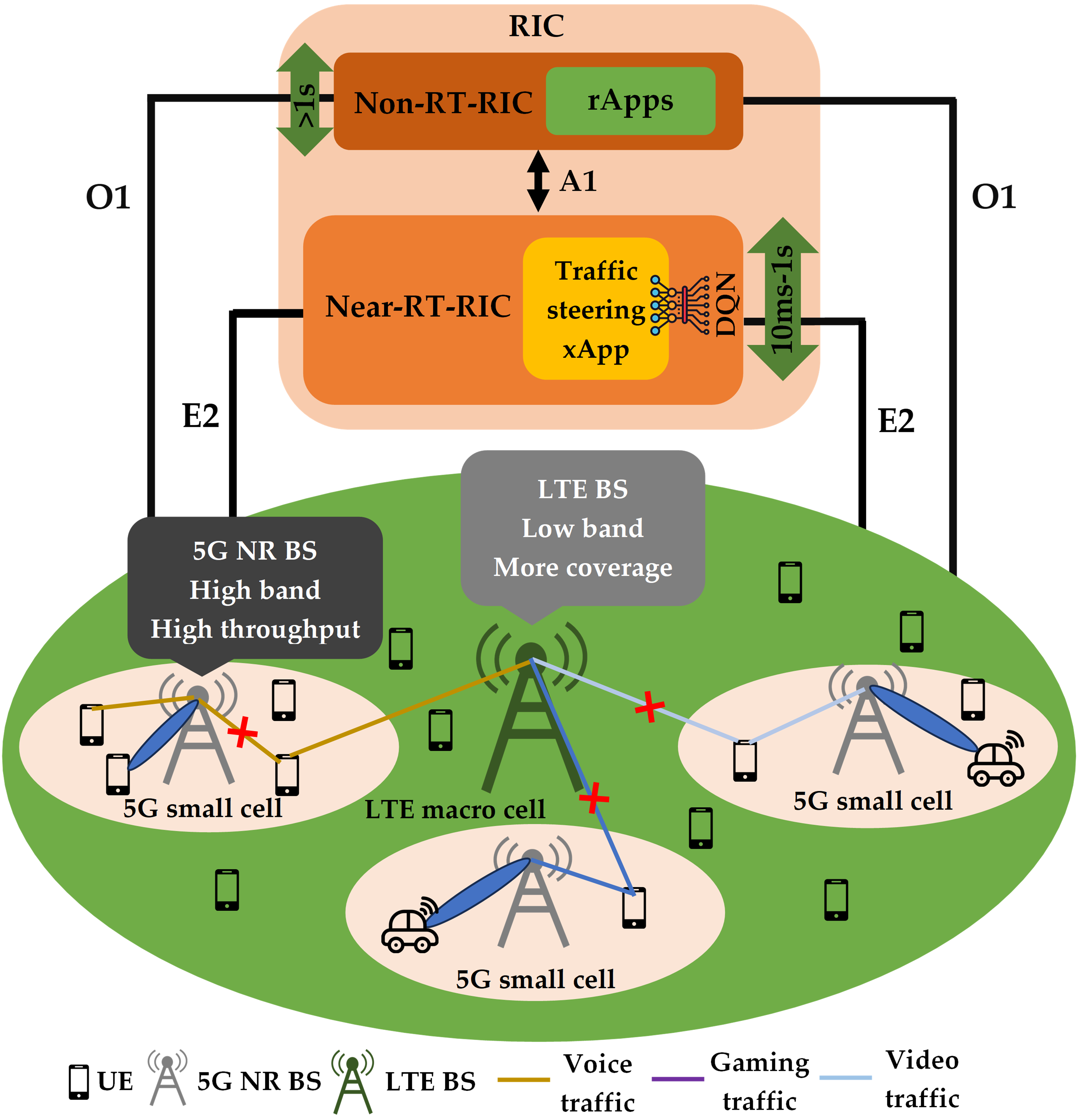}}
\caption{O-RAN architecture with traffic steering xApp.}
\label{fig1}
\vspace{-20pt}
\end{figure}

\vspace{-5pt}
\section{Traffic Steering in O-RAN}
\label{s2}
Traffic steering directs the traffic to appropriate RAT connections for User Equipment (UE) within a mobile network. In multi-RAT scenarios such as the coexistence of 4G, 5G, and Wi-Fi, decisions must be made regarding which cell, RAT, and band a UE should be connected to. Once the BS-UE associations are determined, various traffic types such as voice traffic or video traffic may be handled by different RATs depending on their quality-of-service requirements. 

Conventional traffic steering schemes involve creating a rule-based or deterministic approach to manage traffic in a mobile network. An example of such a traffic steering scheme can be found in \cite{17}, which employs a predetermined threshold by considering the BS load, channel conditions, and the type of user service. However, threshold-based techniques fail to consider that certain UEs may have unique requirements regarding their traffic types. Steering their traffic to an appropriate RAT could lead to a performance increase. As a result, traffic steering becomes a crucial technology that can aid in allocating connections to different RATs to maximize network efficiency, accompanied by stringent maintenance of quality-of-service. 

Meanwhile, the emergence of O-RAN has led to intelligent traffic steering solutions based on advanced ML algorithms, and RIC plays a prime role in such solutions. The non-RT-RIC serves as a software platform for rApps for high-level RAN optimization. It can utilize historical data, traffic patterns, and machine learning algorithms to perform network optimization in non-real-time ($>1s$). On the other hand, near-RT-RIC hosts software applications named xApps to control RAN elements and functions. It enables optimization of the RAN elements and functions in near real-time ($\leq 1ms$). Then, we can execute control signals from xApps to further enhance control over the BS operations. Non-RT-RIC and near-RT-RIC can connect using an A1 interface that works as an information exchange hub. Especially, it can be used to provide ML-based feedback to near-RT-RIC by the non-RT-RIC. Consider a DQN-based traffic steering as an example, one can use the network parameters such as traffic type and BS load level to define environment states. The traffic steering xApp within the near-RT-RIC can employ its integrated DQN algorithm \cite{7} to determine the optimal traffic steering decisions by considering the observed states and optimization goals. Finally, these action policies are transmitted to execute operations in BSs. Fig. \ref{fig1} elaborates on the O-RAN architecture with the traffic steering xApp.

\section{ML-enabled Traffic Steering Schemes}
\label{s3}

This section reviews the latest ML techniques and their applications for traffic steering in O-RAN. Initially, we discuss ML methods that have demonstrated effectiveness in devising efficient traffic steering solutions based on the existing literature. Subsequently, we explore the potential application of unutilized ML algorithms, providing insights into their use for developing resilient traffic steering solutions, which, to date, have not been explored in the literature. Lastly, Table \ref{tab1} summarizes the ML algorithms used for traffic steering, their main features, challenges, and typical applications.

\subsection {State-of-the-art ML Approaches for Traffic Steering}
\subsubsection{Deep Learning}
Multiple input features derived from the contemporary network states can be used to produce predictions on traffic steering using deep learning. This can help to achieve desired objectives, such as load balancing or congestion mitigation. An example of such work is proposed by Fatemeh et al. in which a Long Short-Term Memory (LSTM) network has been used to learn network traffic patterns. The network can predict unknown incoming traffic packets from the network and perform traffic steering accordingly \cite{13}. Furthermore, Lacava et al. propose a Convolutional Neural Network (CNN) to optimally assign a serving BS to each user in the network to steer their traffic \cite{1}. 

However, deep learning-based traffic steering schemes are dependent on labeled network datasets that may be inaccessible in the real world. Lack of data may cause insufficient training leading to erroneous traffic steering decisions.

\subsubsection{Reinforcement Learning}
RL has been extensively used in optimizing Open RAN related problems. Traffic steering in O-RAN involves managing a dynamic and complex environment with numerous variables such as network congestion, varying traffic patterns, and dynamic user demands. RL can adapt to these dynamic conditions and learn optimal traffic steering strategies. Examples of using RL algorithms for traffic steering applications include Q-learning, Deep RL (DRL), and HRL.
Q-learning can be used to perform traffic steering for simplicity \cite{1}, but a longer convergence time is the biggest issue in tabular Q-learning. Especially, when the state-action space is large, Q-learning can perform poorly. DQN can be utilized to address challenges associated with extensive state-action spaces and tackle the problem of sluggish convergence. Instead of relying on vast Q-tables, deep reinforcement learning employs neural networks to predict state-action values. An example of such implementation for the traffic steering problem can be found in \cite{3}. 

\subsubsection{Federated Learning}
Federated learning can be applied to traffic steering by leveraging the distributed intelligence of UEs or edge devices to collectively optimize traffic management decisions. By utilizing federated learning, UEs can collectively contribute their local insights and training data to improve traffic steering decisions without compromising data privacy \cite{5}. Communication overhead due to parameter exchange of the trained models is a critical issue in federated learning. Additionally, the diversity in system capabilities, including variations in storage, computing power, and communication capacities among local devices, poses a significant challenge when implementing federated learning.

\subsubsection{Hierarchical Learning}
Hierarchical learning allows agents to learn high-level policies for goal-directed behavior while also learning lower-level policies for handling specific subtasks.
From the traffic steering point of view, we can further improve the performance of DQN by integrating hierarchical learning. It can gain performance improvement by decomposing complex tasks like traffic steering into two levels of hierarchy \cite{14}. However, hierarchical learning may involve breaking down complex concepts or tasks into simpler subcomponents and sometimes can lead to oversimplification.

\subsection{Potential ML paradigms for Traffic Steering}
Actor-critic algorithms are popular RL techniques that combine aspects of both value-based and policy-based methods \footnote{In RL, a policy maps states to actions, guiding an agent's behavior to maximize cumulative reward. This policy is learned through interaction with the environment. In O-RAN, policies refer to high-level guidelines from the Non-RT-RIC that influence the Near-RT-RIC and its xApps.}. Deep Deterministic Policy Gradient (DDPG) is an example of an actor-critic algorithm that can handle high dimensional and continuous state space with better exploration efficiency. The actor can represent the policy network that decides how to steer traffic, while the critic can evaluate the actions taken by the actor based on the network's performance \cite{18}.

We can use unsupervised learning techniques like k-means clustering or Density-Based Spatial Clustering of Applications with Noise (DBSCAN) to cluster UEs with similar traffic patterns. Then, traffic of these UEs can be steered to a certain BS that can best suit their quality-of-service requirements. One of the major drawbacks of unsupervised learning is that it requires human interpretation and subjective judgment to make sense of the discovered patterns or clusters. Such issues may degrade the expected automation in the Open RAN paradigm since a human expert is needed in the background to make traffic steering decisions. This is where hierarchical learning can play a role. We can put a clustering technique on top of the hierarchy and use the clustering information in the lower-level RL-based controller to make autonomous decisions. This alleviates the need for human intervention.  

Graph learning is another potential ML solution for traffic steering problems. Specifically, Graph Neural Networks (GNNs) can be a suitable candidate to perform traffic steering in O-RAN. Cells and UEs can be used as nodes and the wireless links can be considered as edges. We can perform node classification on the labels of the wireless nodes to sort out the best possible BS-UE connection for traffic steering \cite{20}. Despite the huge potential, graph-based traffic steering solutions may face a great challenge when there is high mobility in the system. 
We can also consider transfer learning as a suitable ML technique for designing traffic steering algorithms for O-RAN. The pre-trained model in transfer learning can guide the decision-making process by providing insights and strategies for efficient traffic steering based on prior learning.

The algorithms presented in this subsection can be adapted to a hierarchical learning setup within an O-RAN architecture. The top layer may employ unsupervised learning algorithms to cluster user equipment based on traffic patterns, providing strategic guidance for traffic distribution. The middle layer can utilize GNNs to continuously update the network graph, reflecting current conditions and aiding in real-time decision-making. At the bottom layer, actor-critic algorithms can make detailed traffic steering decisions, informed by insights from the upper layers and real-time network performance data. Spanning across these layers is transfer learning, which will share knowledge and insights, ensuring adaptability and informed decision-making at all levels, leading to a more efficient, adaptable, and automated traffic management process in the O-RAN system.  

\begin{table*}[!t]
\caption{ML Techniques For Traffic Steering in O-RAN}
\centering
\begin{tabular}{|c|c|c|c|c|}
\hline
\begin{tabular}[c]{@{}c@{}}Learning\\    methods\end{tabular} &
  \begin{tabular}[c]{@{}c@{}}Typical\\    algorithms\end{tabular} &
  Main features &
  Challenges &
  Applications \\ \hline
\begin{tabular}[c]{@{}c@{}}Deep\\    learning\end{tabular} &
  \begin{tabular}[c]{@{}c@{}}CNN \cite{1} and \\ LSTM \cite{13}\end{tabular} &
  \begin{tabular}[c]{@{}c@{}}Maps inputs to outputs\\ within a given dataset, necessitating \\ the availability of labeled data.\end{tabular} &
  \begin{tabular}[c]{@{}c@{}}Dependent on\\  the availability of datasets. \\ Time consuming network training.\end{tabular} &
  \begin{tabular}[c]{@{}c@{}}Traffic congestion \\ prediction to perform\\ traffic steering.\end{tabular} \\ \hline
\multirow{3}{*}{\begin{tabular}[c]{@{}c@{}}Reinforcement\\ learning\end{tabular}} &
  Q-learning \cite{1} &
  \begin{tabular}[c]{@{}c@{}}Aims to find the optimal \\ action-selection policy for an agent \\ via Markov decision processes.\end{tabular} &
  \begin{tabular}[c]{@{}c@{}}Problems having large \\ state-action space require\\ a long time to converge.\end{tabular} &
  \multirow{3}{*}{\begin{tabular}[c]{@{}c@{}}Maximizing network \\ throughput, minimizing \\ network delay,\\ and reducing packet\\ drop rate \\ in a multi-RAT \\ environment\\ with multiple traffic \\ types.\end{tabular}} \\ \cline{2-4}
 &
  DRL \cite{6,3} &
  \begin{tabular}[c]{@{}c@{}}Instead of using Q-tables, \\ DRL combines neural networks \\ with the RL framework to\\  predict state-action values. \\ Therefore, DRL algorithms can \\ handle large state-action spaces.\end{tabular} &
  \begin{tabular}[c]{@{}c@{}}Time-consuming network training, \\ tedious hyperparameter tuning, \\ network training instability, \\ and low sample efficiency.\end{tabular} &
   \\ \cline{2-4}
 &
  HRL \cite{14} &
  \begin{tabular}[c]{@{}c@{}}Decomposes problem into \\ two levels of hierarchy for a better \\ exploration and faster convergence. \\ Handles long-time sparse feedback.\end{tabular} &
  \begin{tabular}[c]{@{}c@{}}Specific to certain problems.\\  Absence of generalization.\end{tabular} &
   \\ \hline
\begin{tabular}[c]{@{}c@{}}Federated\\ learning\end{tabular} &
  \begin{tabular}[c]{@{}c@{}}Federated\\  DRL, \\ Federated \\ meta-learning \cite{5}\end{tabular} &
  \begin{tabular}[c]{@{}c@{}}Involves training algorithms \\ on distributed local datasets \\ without the need to share the \\ actual training data. \\ Handles data with privacy.\end{tabular} &
  \begin{tabular}[c]{@{}c@{}}Hard to handle device\\ heterogeneity and communication \\ overhead. Training a global model\\  across multiple UEs can be \\ complex for non-identically\\  distributed data.\end{tabular} &
  \begin{tabular}[c]{@{}c@{}}UE-centric traffic \\ steering to improve \\ system performance \\ and security.\end{tabular} \\ \hline
\begin{tabular}[c]{@{}c@{}}Hierarchical \\ learning\end{tabular} &
  HRL \cite{14} &
  \begin{tabular}[c]{@{}c@{}}Involves breaking down \\ complex tasks or concepts into\\  smaller, more manageable subtasks.\end{tabular} &
  \begin{tabular}[c]{@{}c@{}}May lead to oversimplification \\ of tasks. Imposes latency, and \\ scalability challenges that \\ need to be carefully managed.\end{tabular} &
  \begin{tabular}[c]{@{}c@{}}Traffic steering\\  based on \\ threshold-based \\ load balancing.\end{tabular} \\ \hline
\end{tabular}
\label{tab1}
\vspace{-15pt}
\end{table*}

\section{Hierarchical Learning Based Traffic Steering}
\label{s4}

\subsection{Hierarchical Learning Scheme}
\begin{figure*}[!t]
\setlength{\abovecaptionskip}{-10pt}
\centerline{\includegraphics[width=1\linewidth]{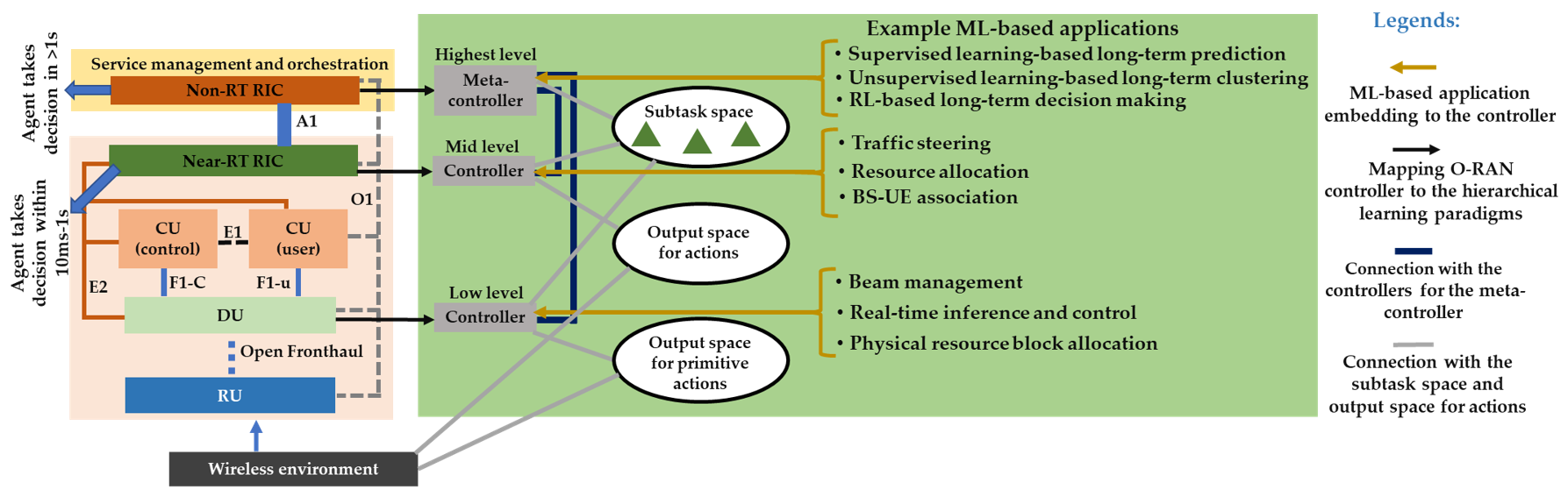}}
\caption{Hierarchical learning scheme for O-RAN.}
\vspace{-15pt}
\label{fig2}
\end{figure*}

Hierarchical learning algorithms can be more efficient when dealing with large and complex network environments. By breaking down the problem into smaller subproblems or subsets, hierarchical algorithms focus on local patterns and relationships at each level. Learning through multiple levels of abstraction reduces the need for extensive exploration and training. As presented in Fig \ref{fig2}, there can be multiple types of algorithms introduced in different levels working in both non-RT-RIC and near-RT-RIC. These algorithms can be embedded as xApps or rApps in RIC. For example, one can combine the principles of both supervised learning and RL within a hierarchical structure. This approach allows for the learning of hierarchical policies that can handle complex tasks by leveraging the advantages of both learning paradigms. At each level of the hierarchy, a supervised learning approach can be used to learn a policy or model that maps the input features to the corresponding actions or decisions specific to that level. Once the initial policies are learned at each level, RL can be used to refine and optimize the policies. The agent engages with the environment, obtains feedback in the shape of rewards or penalties, and refines its policies using techniques such as Q-learning or policy gradient methods. Similarly, we can combine clustering techniques from unsupervised learning with RL. This approach allows an agent to learn and discover the underlying structure of the data and environment while simultaneously optimizing its behavior through RL. 

One of the appealing attributes of hierarchical learning is that we can use different intelligent ML algorithms in non-RT-RIC and near-RT-RIC. Long-term policies and predictions can be determined in the non-RT-RIC by using a wide variety of ML algorithms. 
Using the information or knowledge gathered from the higher level, near-RT-RIC can perform actions in a shorter time scale using ML further. To be exact, an AI agent in non-RT-RIC can make decisions in a time frame greater than 1s. On the other hand, the controller operates at a finer timescale. It makes decisions more frequently, selecting actions at every time step (10ms-1s). This framework is highly suitable for developing applications such as rApps and xApps for O-RAN and allows interactions between them. 
Apart from this, another advantage of hierarchical learning for the development of the RIC applications for O-RAN is the fact that we can handle long-horizon tasks. 
Using the hierarchical learning scheme, we can design RL algorithms in a hierarchical manner that can handle sparse and long-term feedback from the system. It is worth mentioning that non-RT-RIC and near-RT-RIC in this case may have two different agents with two distinct MDPs. They can work in two different time scales as mentioned before.

The difficulty of online learning from untrained models in O-RAN can be mitigated in several ways. For instance, pre-training models using historical data or simulated environments that allow the hierarchical learning system to commence with informed policies based on captured network dynamics can be used. In addition, a model versioning and update strategy can prove to be valuable since it can periodically update models offline, adhering to O-RAN's restriction on online learning from untrained models.

\vspace{-5pt}
\subsection{Hierarchical Deep-Q-Network based Traffic Steering Apps}
In this sub-section, we propose an h-DQN scheme, a hierarchical learning algorithm that uses DQNs embedded in meta-controller and controller for hierarchical and intelligent decision-making. 

\begin{figure*}[!t]
\centerline{\includegraphics[width=0.90\linewidth]{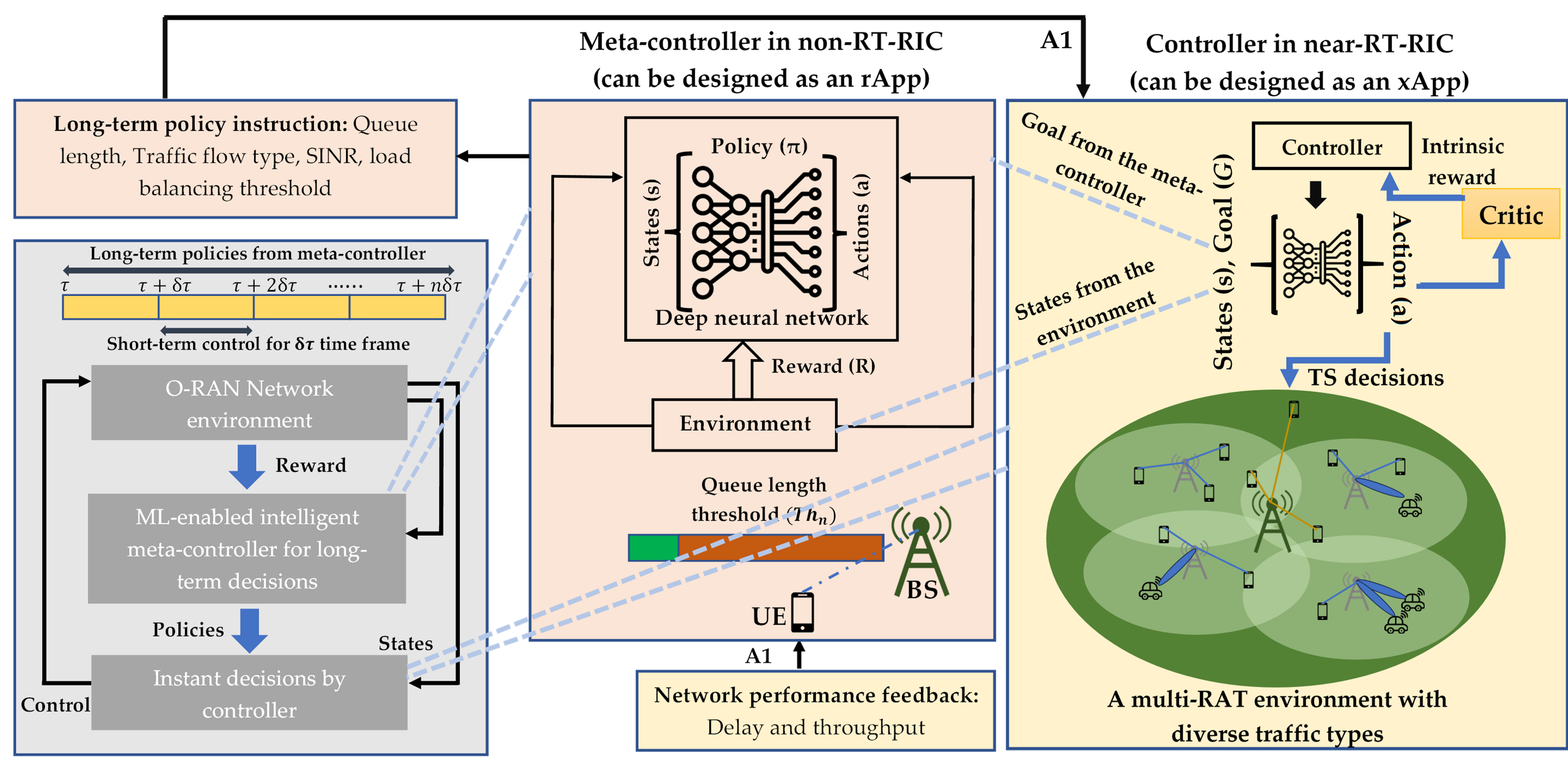}}
\caption{Hierarchical Reinforcement Learning Architecture for Traffic Steering: hierarchical deep Q-network.}
\label{fig3}
\vspace{-15pt}
\end{figure*}

The proposed h-DQN architecture for traffic steering in O-RAN is presented in Fig. \ref{fig3}. The meta-controller located in the non-RT-RIC observes the network environment and can receive a state via an interface that is available depending on the implementation. It sets the goals for the controller based on this observation. However, it does not decide the state or provide it to the controller, which is located inside near-RT-RIC. The states can be network parameters or any other data relevant to conducting traffic steering. For example, if we plan to conduct RAT-specific traffic steering based on UE traffic types, then traffic flow types may be defined as states. Similarly, to perform load balancing via traffic steering, one can consider the load level at each BS as states. Once we define the network parameters to acquire from the environment as states, the meta-controller is supposed to choose a goal. Goals can be set to achieve appropriate load balancing thresholds or to meet a certain quality-of-service requirement. Under high traffic, keeping queue lengths minimal is important to avoid excessive delay before re-transmission or discarding data packets. Conversely, to maximize transmission opportunities, the MAC scheduler should fill the Downlink Shared Channel (DL-SCH) with as much data as the PHY entity requests each Transmission Time Interval (TTI). This approach, which aims to keep buffers full, contradicts the need for shorter queues. To navigate these conflicting requirements, it's essential to find a balance. Sorting out an appropriate threshold for enqueued data packets helps steer traffic effectively, ensuring quality-of-service in terms of delay and optimal throughput.

In the lower-level setup, the controller, which is integrated as an xApp (an RL agent), employs both the goal and state received from the environment to make traffic steering decisions until it either accomplishes the goal or the episode ends. There's an internal critic responsible for assessing goal achievement and assigning suitable rewards to the controller. It interacts directly with the environment by executing actions and observing the resulting states and rewards. It receives the current state directly from the environment (just like any RL agent) and uses the goal set by the meta-controller to guide its actions. The controller aims to maximize the accumulation of intrinsic rewards. The reward mechanism can be generalized to optimize other key performance indicators such as delay or throughput. 

To summarize, there are two agents in this framework. One resides in non-RT-RIC as a meta-controller and another in the near-RT-RIC as a controller. The decision timescales of the meta-controller and the controller are different. The meta-controller operates at a coarser timescale. It makes decisions less frequently, setting high-level goals that guide the overall behavior of the agent for extended periods. The controller operates at a finer timescale. It makes decisions more frequently, selecting actions at every time step to achieve the current goal set by the meta-controller. It can take multiple actions to achieve a goal. The meta-controller will wait until it achieves the goal or reaches a terminal state before making the next decision.

\subsection{OTIC Integration and O-RAN AI/ML Workflow:}

Hierarchical learning-based traffic steering schemes can be integrated and tested with the Open Test and Integration Center (OTIC) approved by the O-RAN alliance \cite{2}. In terms of AI workflows, the bi-level architecture of h-DQN can be integrated, allowing lower-level agents (xApps) to make immediate decisions and higher-level agents (rApps) to adjust global strategies.

Furthermore, it is also compatible with AI/ML workflow description and requirements presented by the O-RAN Alliance group \cite{21}. Data from RAN components and the RIC are collected via the O1 interface to a data collector at the Service Management and Orchestration (SMO), potentially using a data bus like Kafka \cite{13}. An rApp utilizes this data to provide state and goal inputs for optimizing delay and throughput based on quality-of-service requirements. The non-RT-RIC can query an ML/AI model within the SMO's AI server, and upon model training, the non-RT-RIC can be notified of the inference. In such an implementation, the traffic steering xApp in the near-RT-RIC can be updated via the A1 interface. It makes traffic steering decisions based on rewards from a reward function, considering specific traffic loads and a learned goal like a load balancing threshold. 

In addition, we summarize the overall workflow of the proposed hierarchical learning scheme. As mentioned before, we have two different agents. In the proposed system, the actuators refer to the network elements, e.g., eNB and gNBs, that handle the actual traffic flow based on the decisions made by the agents. These network elements implement the traffic steering decisions by managing the traffic data to UEs. In particular, the BSs will collect and report network status such as SINR measurements, queue lengths, and traffic types, to the controllers. Considering the traffic steering example, the controller will send command signals to change the load balancing thresholds, e.g., “set the queue length threshold for 5 Mbps traffic load to 0.8”, and make traffic steering decisions, i.e., “steer voice traffic flow type to RAT 1”.

\section{A Case Study with Hierarchical Reinforcement Learning for 5G}
\label{s5}

This section presents a case study of using HRL for traffic steering in O-RAN. The case study examines a multi-RAT network, where numerous users are linked through dual connectivity to both 5G and LTE RATs. There are small cells equipped with 5G NR BSs operating at 3.5 GHz. These small cells are positioned within the coverage area of a macro-cell, facilitated by a single LTE BS operating at 0.8 GHz. The operational bandwidths for the LTE and 5G NR BSs are 10 MHz and 20 MHz. Traffic flows can either be steered to eNB or gNB based on the decision of the h-DQN agent. This simulation involves a total of 60 users and encompasses three distinct types of traffic: video, gaming, and voice (one traffic type per UE). To ensure quality-of-service, we have established specific requirements for each traffic type based on the 3GPP specifications \cite{14}. We have used 3GPP traffic models \cite{15} for three traffic types. 
The simulation includes a lower-level controller operating under the guidance of goals determined and suggested by a meta-controller. The meta-controller keeps track of the state of the external communication environment and chooses goals accordingly. Both controllers are separate RL agents having own MDPs interacting with the environment. The MDPs of the meta-controller in non-RT-RIC and the controller in near-RT RIC are:

\noindent
$\bullet$~\textit{MDP of the meta-controller:} The state of the meta-controller includes traffic types, the SINR measurements, and the queue length for each type of RAT. The goals for the controller are defined by thresholds related to the queue lengths. These thresholds are used to decide when to defer transmission to another RAT for load-balancing purposes. The extrinsic reward function for the meta-controller is derived from the overall objective of the system, which is calculated as the average of the intrinsic rewards over a specified number of steps.

\noindent
$\bullet$~\textit{MDP of the controller:} The controller shares the same states as the meta-controller. Flow admission to different RATs is considered to be in the action space. We design the intrinsic reward to ensure satisfactory performance, focusing on network delay and average system throughput.

\begin{figure}[!t]
\setlength{\abovecaptionskip}{-5pt}
\centerline{\includegraphics[width=0.8\linewidth]{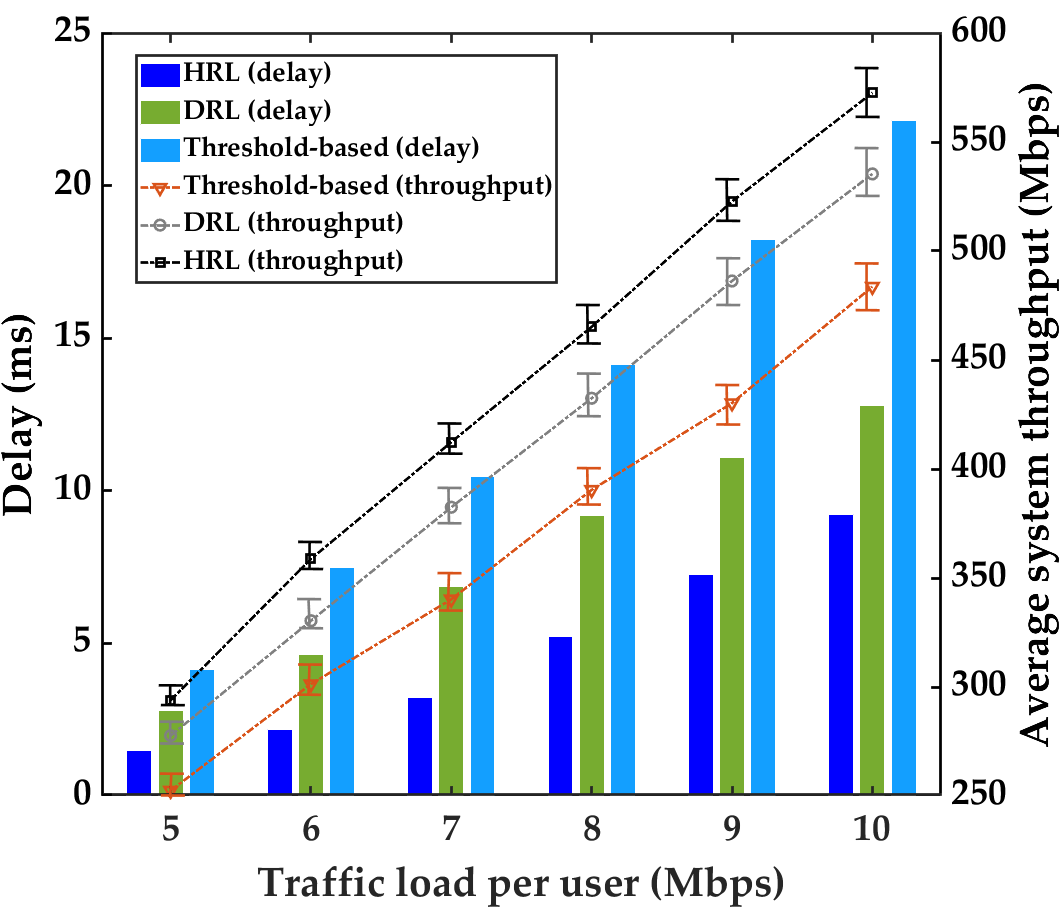}}
\caption{Performance achievement of hierarchical reinforcement learning versus baselines in terms of throughput and delay.}
\label{fig4}
\vspace{-10pt}
\end{figure}

Fig. \ref{fig4} illustrates the comparison between the traffic steering approach based on h-DQN and the baseline algorithms, focusing on system throughput and network delay. Two baseline schemes are considered: one being a threshold-based heuristic algorithm \cite{3}. 
This threshold, denoted as $T_{th}$, is derived by averaging all the aforementioned metrics. Additionally, a variable $W$  is calculated using the same factors, but with an emphasis on weighted metrics. The decision for traffic steering is made by comparing the values of $W$ and $T_{th}$. The second one is based on a DQN algorithm \cite{17}. DQN is chosen as a baseline due to its demonstrated success in reinforcement learning and extensive study in wireless communication applications \cite{3,6,7}. 
The proposed method surpasses both the threshold-based heuristic and the DRL baseline, achieving an average increase in throughput of 15.55\% and 6.46\%, respectively. Furthermore, the h-DQN scheme demonstrates a lower network delay of 27.74\% and 58.96\% compared to the same baselines. The DRL algorithm lacks a dedicated mechanism for adapting to shifts in traffic load, which leads to lower system throughput and higher delay.

\begin{figure}[!t]
\setlength{\abovecaptionskip}{-5pt}
\centerline{\includegraphics[width=0.8\linewidth]{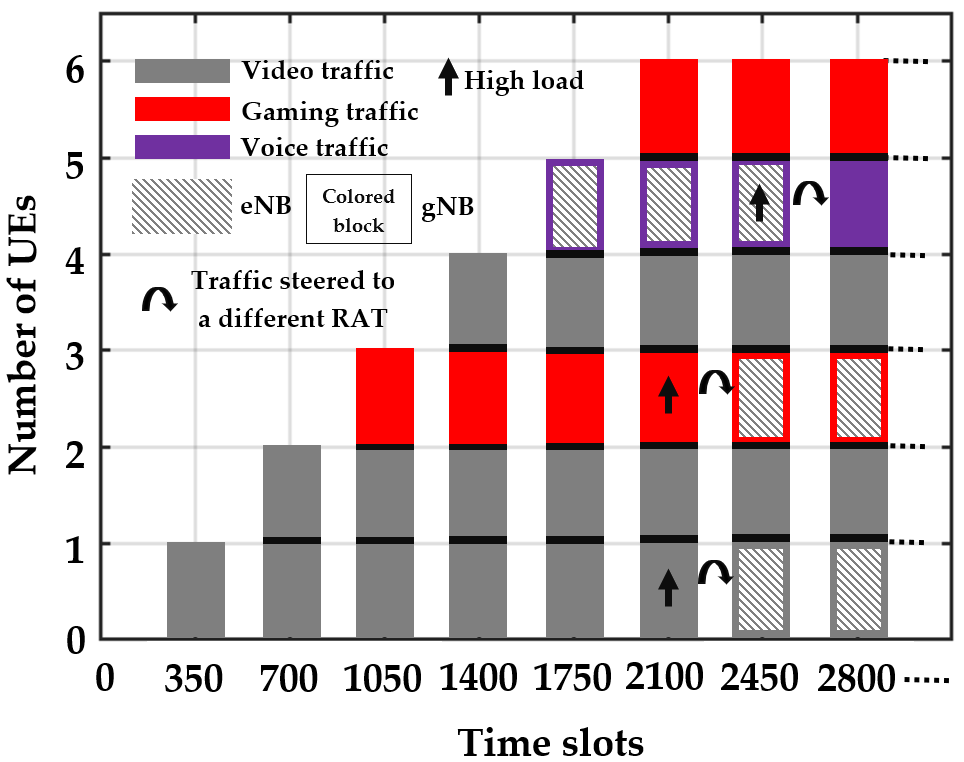}}
\caption{Traffic being steered to a different RAT based on load balancing threshold.} 
\label{fig5}
\vspace{-18pt}
\end{figure}

To illustrate how traffic steering gets performed by the traffic steering xApp, we increase the number of UEs by one after some time slots and observe which BS the traffic is getting steered to. We use queue length to indicate the load at each BS. Whenever a high load is experienced because too many UEs are getting served by a single small cell BS, we can observe the traffic getting steered to a different RAT. Fig. \ref{fig5} illustrates the traffic steering between different RATs when a BS experiences a high load and the load balancing threshold is exceeded. The grey, red, and blue colors represent the video, gaming, and voice traffic respectively. 
In the scenario described, a heavy load occurs when a higher number of UEs are simultaneously served by the same BS in the small cell during the $2100^{th}$ time slot. Consequently, the data traffic of the fifth UE is redirected to another RAT during the $2450^{th}$ time slot. A similar situation occurs for the third UE, with its traffic being steered to a different RAT during the $2100^{th}$ time slot. 


\section{Conclusions}
\label{s6}
ML techniques offer promising opportunities for intelligent traffic steering in O-RAN. In this work, we first provided background on traffic steering in O-RAN and surveyed related ML techniques that can be used in wireless networks for traffic steering. Then, we presented an HRL framework for O-RAN and how h-DQN can be a robust and suitable candidate for traffic steering in such an environment. The proposed h-DQN-based traffic steering scheme gains a significant performance increase based on our simulation results compared to the DRL and heuristic baselines. We plan to incorporate evaluations using O-RAN testbeds in our future work as well. This will allow us to validate our findings under real-world conditions and further enhance the robustness and applicability of our approach.  

\section*{Acknowledgment}
\vspace{-5pt}
This work has been supported by MITACS and Ericsson Canada, and NSERC Canada Research Chairs Program. 
\vspace{-5pt}

\bibliographystyle{IEEEtran}
\bibliography{Reference}

\begin{thebibliography}{10}
\providecommand{\url}[1]{#1}
\csname url@samestyle\endcsname
\providecommand{\newblock}{\relax}
\providecommand{\bibinfo}[2]{#2}
\providecommand{\BIBentrySTDinterwordspacing}{\spaceskip=0pt\relax}
\providecommand{\BIBentryALTinterwordstretchfactor}{4}
\providecommand{\BIBentryALTinterwordspacing}{\spaceskip=\fontdimen2\font plus
\BIBentryALTinterwordstretchfactor\fontdimen3\font minus
  \fontdimen4\font\relax}
\providecommand{\BIBforeignlanguage}[2]{{%
\expandafter\ifx\csname l@#1\endcsname\relax
\typeout{** WARNING: IEEEtran.bst: No hyphenation pattern has been}%
\typeout{** loaded for the language `#1'. Using the pattern for}%
\typeout{** the default language instead.}%
\else
\language=\csname l@#1\endcsname
\fi
#2}}
\providecommand{\BIBdecl}{\relax}
\BIBdecl

\bibitem{2}
M.~Polese, L.~Bonati, S.~D'Oro, S.~Basagni, and T.~Melodia, ``Colo-ran:
  Developing machine learning-based xapps for open ran closed-loop control on
  programmable experimental platforms,'' \emph{IEEE Transactions on Mobile
  Computing}, vol.~22, no.~10, pp. 5787--5800, 2023.

\bibitem{21}
O.~Alliance, ``{O-RAN Working Group 2 AI/ML Workflow Description and
  Requirements},'' \emph{ORAN-WG2. AIML. v01}, vol.~1, 2019.

\bibitem{4}
\BIBentryALTinterwordspacing
M.~Dryjanski, L.~Kulacz, and A.~Kliks, ``Toward {M}odular and {F}lexible {O}pen
  {RAN} {I}mplementations in {6G} {N}etworks: {T}raffic {S}teering {U}se {C}ase
  and {O-RAN} x{A}pps,'' \emph{Sensors}, vol.~21, no.~24, 2021. [Online].
  Available: \url{https://www.mdpi.com/1424-8220/21/24/8173}
\BIBentrySTDinterwordspacing

\bibitem{5}
H.~Erdol, X.~Wang, P.~Li, J.~D. Thomas, R.~Piechocki, G.~Oikonomou, R.~Inacio,
  A.~Ahmad, K.~Briggs, and S.~Kapoor, ``Federated {M}eta-{L}earning for
  {T}raffic {S}teering in {O-RAN},'' in \emph{2022 IEEE 96th Vehicular
  Technology Conference (VTC2022-Fall)}, 2022, pp. 1--7.

\bibitem{1}
A.~Lacava, M.~Polese, R.~Sivaraj, R.~Soundrarajan, B.~S. Bhati, T.~Singh,
  T.~Zugno, F.~Cuomo, and T.~Melodia, ``Programmable and {C}ustomized
  {I}ntelligence for {T}raffic {S}teering in 5{G} {N}etworks {U}sing {O}pen
  {RAN} {A}rchitectures,'' \emph{IEEE Transactions on Mobile Computing}, pp.
  1--16, 2023.

\bibitem{6}
I.~Tamim, S.~Aleyadeh, and A.~Shami, ``Intelligent {O-RAN} {T}raffic {S}teering
  for {URLLC} {T}hrough {D}eep {R}einforcement {L}earning,'' in \emph{ICC 2023
  - IEEE International Conference on Communications}, 2023, pp. 112--118.

\bibitem{12}
\BIBentryALTinterwordspacing
T.~D. Kulkarni, K.~Narasimhan, A.~Saeedi, and J.~B. Tenenbaum, ``Hierarchical
  {D}eep {R}einforcement {L}earning: {I}ntegrating {T}emporal {A}bstraction and
  {I}ntrinsic {M}otivation,'' \emph{CoRR}, vol. abs/1604.06057, 2016. [Online].
  Available: \url{http://arxiv.org/abs/1604.06057}
\BIBentrySTDinterwordspacing

\bibitem{14}
\BIBentryALTinterwordspacing
M.~A. Habib, H.~Zhou, P.~E. Iturria-Rivera, M.~Elsayed, M.~Bavand, R.~Gaigalas,
  Y.~Ozcan, and M.~Erol-Kantarci, ``Hierarchical {R}einforcement {L}earning
  {B}ased {T}raffic {S}teering in {M}ulti-{RAT} {5G} {D}eployments,'' Jan.
  2023. [Online]. Available: \url{https://arxiv.org/abs/2301.07818}
\BIBentrySTDinterwordspacing

\bibitem{3}
M.~A. Habib, H.~Zhou, P.~E. Iturria-Rivera, M.~Elsayed, M.~Bavand, R.~Gaigalas,
  S.~Furr, and M.~Erol-Kantarci, ``Traffic {S}teering for 5{G} {M}ulti-{RAT}
  {D}eployments using {D}eep {R}einforcement {L}earning,'' in \emph{2023 IEEE
  20th Consumer Communications \& Networking Conference (CCNC)}, 2023, pp.
  164--169.

\bibitem{17}
M.~Khaturia, P.~Jha, and A.~Karandikar, ``5{G}-{F}low: A {u}nified
  {M}ulti-{RAT} {RAN} {a}rchitecture for {b}eyond 5{G} {n}etworks,''
  \emph{Computer Networks}, vol. 198, p. 108412, 2021.

\bibitem{7}
V.~Mnih, K.~Kavukcuoglu, D.~Silver, A.~A. Rusu, J.~Veness, M.~G. Bellemare,
  A.~Graves, M.~Riedmiller, A.~K. Fidjeland, G.~Ostrovski \emph{et~al.},
  ``Human-{L}evel {C}ontrol {T}hrough {D}eep {R}einforcement {L}earning,''
  \emph{nature}, vol. 518, no. 7540, pp. 529--533, 2015.

\bibitem{13}
F.~Kavehmadavani, V.-D. Nguyen, T.~X. Vu, and S.~Chatzinotas, ``Intelligent
  {T}raffic {S}teering in {B}eyond {5G} {O}pen {RAN} based on {LSTM} {T}raffic
  {P}rediction,'' \emph{IEEE Transactions on Wireless Communications}, pp.
  1--1, 2023.

\bibitem{18}
\BIBentryALTinterwordspacing
T.~P. Lillicrap, J.~J. Hunt, A.~Pritzel, N.~Heess, T.~Erez, Y.~Tassa,
  D.~Silver, and D.~Wierstra, ``{C}ontinuous {C}ontrol with {D}eep
  {R}einforcement {L}earning,'' Sep. 2015. [Online]. Available:
  \url{https://arxiv.org/abs/1509.02971}
\BIBentrySTDinterwordspacing

\bibitem{20}
O.~Orhan, V.~N. Swamy, T.~Tetzlaff, M.~Nassar, H.~Nikopour, and S.~Talwar,
  ``Connection {M}anagement x{APP} for {O-RAN RIC}: A {G}raph {N}eural
  {N}etwork and {R}einforcement {L}earning {A}pproach,'' in \emph{2021 20th
  IEEE International Conference on Machine Learning and Applications (ICMLA)},
  2021, pp. 936--941.

\bibitem{15}
J.~Navarro-Ortiz, P.~Romero-Diaz, S.~Sendra, P.~Ameigeiras, J.~J. Ramos-Munoz,
  and J.~M. Lopez-Soler, ``A {S}urvey on 5{G} {U}sage {S}cenarios and {T}raffic
  {M}odels,'' \emph{IEEE Communications Surveys \& Tutorials}, vol.~22, no.~2,
  pp. 905--929, 2020.

\end{thebibliography}

\end{document}